\documentclass[referee]{aa}
\usepackage{txfonts}
\usepackage[dvips]{graphicx}

\newcommand{\gtappeq}{\raisebox{-0.6ex}{$\,\stackrel
{\raisebox{-.2ex}{$\textstyle >$}}{\sim}\,$}}

\begin{document}

\title{Determining the cosmic ray ionization rate in dynamically evolving 
clouds}
\author{C.J. Lintott \and J.M.C. Rawlings} 

\institute{Department of Physics and Astronomy, University College London, Gower 
Street, London, WC1E 6BT, UK\\ 
\email{cjl@star.ucl.ac.uk}}

\date{Received <date> / Accepted <date>}

\abstract{
The ionization fraction is an important factor in determining the chemical and 
physical evolution of star forming regions. In the dense, dark starless cores 
of such objects, the ionization rate is dominated by cosmic rays; it is 
therefore possible to use simple analytic estimators, based on the relative 
abundances of different molecular tracers, to determine the cosmic ray 
ionization rate.
This paper uses a simple model to investigate the accuracy of two well-known
estimators in dynamically evolving molecular clouds.
It is found that, although the analytical formulae based on the abundances of $\mathrm{H_3^+}$,$\mathrm{H_2}$,CO,O,$\mathrm{H_2O}$ and $\mathrm{HCO^+}$) give a reasonably accurate
measure of the cosmic ray ionization rate in static, quiescent clouds, 
significant discrepancies occur in rapidly evolving (collapsing) clouds.
As recent evidence suggests that molecular clouds may consist of complex, 
dynamically evolving sub-structure, we conclude that simple abundance ratios
do not provide reliable estimates of the cosmic ray ionization rate in 
dynamically active regions.

\keywords{
astrochemistry -- stars : formation}

}

\authorrunning{Lintott \& Rawlings}
\titlerunning{Cosmic Ray Ionization Rate in Dynamically Evolving Clouds}

\maketitle

\section{INTRODUCTION}

In this paper we examine the accuracy of the methods used to constrain the
rate of cosmic-ray induced ionization in molecular clouds. 
 The level of ionization within a dark cloud may determine whether or not the {cloud is stable against gravitational 
collapse leading to the formation of proto-stellar objects. This is particularly true if magnetic 
fields permeate the cloud - in which case the process is controlled by the 
coupling between the ions (tied to the magnetic field lines) and the neutral species. 
This applies both to quasi-static contraction (via ambipolar diffusion) and 
turbulent dissipation.
Ruffle et al. (\cite{rufflehartquist}), investigated these processes and showed that the stability of low extinction 
(A$_v<3$) clumps in the Rosette Molecular Cloud could be explained in terms of 
photochemically driven high ionization levels, which inhibit the damping of 
(MHD) turbulent support.

Cosmic rays also play a very important r\^ole in the physics of the interstellar 
medium - for example, they are an important source of heating in dark molecular clouds (e.g. 
Goldsmith \& Langer \cite{goldsmithlanger}). These dark clouds are highly opaque to optical and ultraviolet (ionizing) 
interstellar radiation, yet they are known to contain a wide variety of atomic 
and molecular ions.
Although dust grain surface chemistry is extremely important in dark and 
diffuse clouds (most significantly in the synthesis of H$_2$) 
the chemistry within dark clouds is driven by gas-phase 
reactions and the most significant of these (particularly for oxygen and nitrogen-bearing species) are ion-neutral reactions
.
The source of this ionization is believed to be cosmic rays. In the case of 
diffuse media the initiating reaction is 

\[ {\rm H + crp \longrightarrow H^+ + e^-} \] 

(where crp stands for `cosmic ray particle')

whilst in dense molecular clouds approximately 97\% of cosmic ray-H$_2$ impacts lead 
to the formation of H$_2^+$ (Hartquist \& Williams 1995) via the reaction:
\begin{equation} \label{reac1}
{\rm H_2 + crp \longrightarrow H_2^+ + e^-} 
\end{equation}
which is followed by the rapid reaction:
\begin{equation} \label{reac2}
{\rm H_2^+ + H_2 \longrightarrow H_3^+ + H}
\end{equation}
Subsequently, the H$_3^+$ then drives ion-molecule chemistries (see below).
The rate of ionization and electron abundance within these clouds therefore has
enormous implications for the efficiency of the molecular chemistry and the
overall physical stability of many dark clouds.

As a result of these reactions, 
nearly every cosmic-ray ionization creates an H$_3^+$ ion and 
this reacts with an abundant metal, such as oxygen or nitrogen (see below).
The rate coefficient for reaction (\ref{reac1}), $\zeta$, is therefore a key 
parameter in models of dark cloud chemistry;
the relevant timescale for a cosmic-ray ionization induced chemistry to 
reach equilibrium is of the order of the fractional abundance of the metals
(approximately 10$^{-4}$) divided by $\zeta$. 
A value of $\zeta = 1.3\times 10^{-17}$s$^{-1}$ is often quoted in the 
literature, although in most cases only an order 
of magnitude accuracy is appropriate. In some sources, much higher values of 
$\zeta$ have been inferred (e.g. at least $\zeta = 8\times 10^{-16}$s$^{-1}$ in 
$o$ Per, van Dishoeck \& Black \cite{vandishoeckblack}).
Generally, studies typically indicate that $\zeta=10^{-16}-10^{-17}$s$^{-1}$
per hydrogen atom. 

However, if we assume a `standard' value of $\zeta\sim 10^{-17}$s$^{-1}$, the
chemical timescale is approximately 10$^6$ years. 
Usually, this assumption of equilibrium is adopted when using molecular 
abundances to infer the value of $\zeta$, but here we note that
the free-fall collapse timescale of a gas cloud is $3.4\times 10^7 n^{-0.5}$ 
years (Spitzer \cite{spitzer}) so, for typical dark cloud densities of 
$10^3-10^4$cm$^{-3}$, a dynamical timescale of $3.4\times 10^5$-$10^6$ years 
is inferred. $\zeta$ is usually estimated from observations of the 
abundances of molecular species interpreted through simple chemical models. 
Such methods depend on the assumption that chemical equilibrium has already 
been reached. It is not obvious, therefore, that estimates of the 
cosmic ray ionization rate in regions which may be rapidly evolving are correct.

The rates of ionization have been clearly established for diffuse clouds, 
largely through observations of OH and HD (see below). In dark clouds the
situation is not so clear. 
The standard assumption is that the same value of $\zeta$ is applicable in dark 
clouds as in diffuse clouds. 
However, even in diffuse clouds, there is now believed to be very considerable 
variation in $\zeta$ from source to source. Thus, in a careful study based on 
H$_3^+$ observations in $\zeta$~Persei, McCall et al. (\cite{mccalletal}), deduce that 
$\zeta$ is as high as $1.2\times 10^{-15}$s$^{-1}$. However, Le Petit et al. (2004) find that their model reproduces observed abundances in the same source using a value of $\zeta = 2.5 \times 10^{-16}$.
In addition, as dark clouds are very much denser 
and also are permeated by ordered magnetic fields there is no {\em a priori} 
reason to assume that $\zeta$ is the same in diffuse and dark clouds.
Moreover, as indicated above, attempts to estimate $\zeta$ in dark clouds are
usually based on estimators that make an assumption of chemical equilibrium.

Most matter in a molecular cloud is essentially sterile, in that stars only 
form from a small fraction of the mass of a cloud (Leisawitz et al. \cite{leisawitzbash}). 
Most of this mass exists in regions of low extinction and is effectively in a 
photon-dominated region (Evans \cite{evans}). Thus, most diffuse clouds are either 
supported against gravitational collapse, or else are transient objects that 
are dispersed before collapse and star-formation can occur. This implies that
dark, dense, clouds form from diffuse clouds and may retain some degree of the chemical
identity of their precursors.

In this paper we assess the implications of the assumption of chemical
equilibrium on the inferred cosmic ray ionization rate in
molecular clouds which are evolving on timescales that are short compared to
the time required for chemical equilibrium to be established.

\section{ESTIMATES OF THE COSMIC RAY IONIZATION RATE, $\zeta$}

The usual method of estimating the cosmic ray ionization rate is to employ a 
semi-analytical, semi-empirical technique. 
Thus what is, in reality, a complex multi-channel chemistry is simplified to to 
a small number of primary loss and formation reactions. The absolute 
abundances, or abundance ratios, of key tracers are then interpreted through 
simple analytical expressions. 
In practise, whilst many of the chemical abundances in these expressions are 
directly measureable (such as those for CO, HCO$^+$ and H$_3^+$), others are 
unobservable (e.g. H$_2$O) or else observational data does not exist.
Consequently, the values used as input to the formulae are a 
combination of observed quantities and predictions from (often complex) 
chemical models.
However, both the derivation of the analytical expressions and the modelled 
abundances used as input almost invariably depend on the assumption of 
chemical equilibrium.
Therefore, for the reasons stated above, it is possible that they may not yield 
accurate values for $\zeta$.    

Lepp, Dalgarno \& Sternberg \cite{leppdalgarno} used the relationship between the abundances 
of (observable) OH and H$_3^+$ and the equilibrium relationship between
n(H$_3^+$) and $\zeta$ (see below) to deduce constraints on $\zeta$ in dense,
dark interstellar clouds. 
This analysis was based on the assumption that the photodissociation of H$_2$O
by cosmic-ray induced ultraviolet photons is a major source of OH in dark
clouds.

In contrast, the abundances of carbon-bearing species in diffuse clouds are not sensitive 
to $\zeta$, and the the best estimators of $\zeta$ come from observations and 
models of the OH and HD abundances (e.g. van Dishoeck \& Black \cite{vandishoeckblack}; Federman 
et al. \cite{federmanweber}). The entry into the oxygen chemistry 
is somewhat different than in dense molecular environments. The source of O$^+$ is cosmic 
ray ionization of hydrogen atoms and resonant charge exchange with oxygen 
atoms;
\[ {\rm H^+ + O \longrightarrow O^+ + H} \]
This is followed by reactions with H$_2$ and dissociative recombination
\[ {\rm O^+ + H_2,e^- \ldots\longrightarrow OH, H_2O} \] 
whilst the HD formation route is 
\[ {\rm H^+ + D \longrightarrow H + D^+} \]
\[ {\rm D^+ + H_2 \longrightarrow HD + H^+} \]
Thus the abundances of both OH and HD are directly proportional to $\zeta$. 
Studies using this reaction network yield a value of $\zeta\sim 10^{-17}-10^{-16}$s$^{-1}$ 
for the rate of cosmic ray ionization of H atoms in diffuse clouds (Federman et 
al. \cite{federmanweber}) - the spread of values being partly attributable to different 
assumptions being made and partly due to genuine variations, which 
may derive from uncertainties in the temperatures and the
assumed radiation fields. 

If localised heating (e.g. through the action of shocks) is present, then the
reaction
\[ {\rm O + H_2 \longrightarrow OH + H} \]
can occur, in which case these methods may overestimate the value of $\zeta$.

An alternative estimator is based on the H$_3^+$ abundances. In diffuse clouds, 
H$_3^+$ is formed by reactions \ref{reac1} and \ref{reac2} and lost by 
dissociative recombination:
\[ {\rm H_3^+ + e^- \longrightarrow H_2 + H} \]
although this approach can be undermined by uncertainties in the value of
the dissociative recombination rate co-efficient and the electron density.
 
In diffuse clouds the main loss channel of simple molecular species is 
photodissociation and the implied timescales are of the order of 300 years, 
significantly less than the dynamical timescales of diffuse clouds. In addition 
the chemistry and rate-coefficients are are well-defined.
For these reasons, the inferred ionization rates are believed to be reasonably robust 
(Hartquist and Williams \cite{hartquistwilliams}). However, McCall et al. 2003 present a direct detection of H$_3^+$ in diffuse clouds towards $\zeta$ Per, and infer a correspondingly high cosmic ray ionization rate, 40 times larger than the `standard' value. However, Le Petit et al. 2004, noting that such a high ionization rate may be inconsistent with other molecular and atomic observations (especially those of OH and HS), explain the observed abundance of $\mathrm{H_3^+}$ via a two-phase model. This example illustrates the difficulty of seperating the ionization rate from effects of structure and, as we discuss below, dynamical evolution).

In dark clouds, the reactive species H$_3^+$ provides important entries into 
the ion-neutral chemistry. H$_3^+$ is usually more easily detected in its deuterated 
form, H$_2$D$^+$, to which it is linked via;
\[ {\rm H_3^+ + HD \leftrightarrow H_2D^+ + H_2} \]
This reaction also leads to the deuteration of molecular ions, such as DCO$^+$, 
which can be used to constrain the fractional ionization (Caselli et al. 1998).

The oxygen chemistry follows a simple route:
\begin{equation} \label{reac3}
{\rm H_3^+ + O \longrightarrow OH^+ + H_2} 
\end{equation}
\begin{equation} \label{reac4}
{\rm OH^+ + H_2 \longrightarrow H_2O^+ + H} 
\end{equation}
\begin{equation} \label{reac5}
{\rm H_2O^+ + H_2 \longrightarrow H_3O^+ + H}
\end{equation}
\begin{equation} \label{reac6}
{\rm H_3O^+ + e^- \longrightarrow OH + 2H, H_2O + H, OH + H_2} 
\end{equation}

Other major loss routes for H$_3^+$ are reaction with water;
\begin{equation} \label{reac7}
{\rm H_3^+ + H_2O \longrightarrow H_3O^+ + H_2} 
\end{equation}
followed by dissociative recombination (\ref{reac6}), and reaction with CO, 
leading to HCO$^+$ formation;
\begin{equation} \label{reac8}
{\rm H_3^+ + CO \longrightarrow HCO^+} 
\end{equation}

The HCO$^+$ is lost through dissociative recombination;
\begin{equation} \label{reac9}
{\rm HCO^+ + e^- \longrightarrow CO + H} 
\end{equation}
reaction with water;
\begin{equation} \label{reac10}
{\rm HCO^+ + H_2O \longrightarrow H_3O^+ + CO} 
\end{equation}
and charge transfer with low ionization potential elements;
\begin{equation} \label{reac11}
{\rm HCO^+ + Na,Mg \longrightarrow Na^+,Mg^+ + H + CO}
\end{equation}

This last reaction is particularly important in determining the ionization 
level of dark clouds. In metal poor dark clouds, HCO$^+$ can be the dominant 
ion. However, the radiative recombination rates for elemental ions, 
such as Na$^+$ and Mg$^+$, are very slow so in undepeleted regions, reaction 
(\ref{reac11}) can effectively result in a higher, stable, ionization level.

Williams et al. (\cite{williamsbergin}) used observations of C$^{18}$O, H$^{13}$CO$^+$ and 
DCO$^+$ towards 23 low-mass cores and a steady-state chemical model to deduce a 
mean value of $\zeta = 5\times 10^{-17}$s$^{-1}$. 
The [DCO$^+$]/[HCO$^+$] ratio, which is relatively insensitive to $\zeta$ was 
used to deduce the ionization fraction. In low mass cores, the ionization can 
be enhanced as a result of the presence of newly-formed low mass stars (which 
are X-ray sources) and the penetration of external UV, if the clouds are 
inhomogeneous.
Bergin et al. (\cite{berginplume}) also used obervations of [DCO$^+$]/[HCO$^+$] 
to infer the fractional ionization within Orion massive dark cores, where 
cosmic rays are believed to be the dominant source of ionization.
Both studies initially considered variations of $\zeta$ within a limited range 
($1-15\times 10^{-17}$s$^{-1}$). This was constrained by the condition of 
thermal balance at the temperatures and densities of the sources being 
considered, recognising that cosmic rays are the dominant source of heating in 
starless cores. They subsequently adopted a value of $\zeta = 5\times 
10^{-17}$s$^{-1}$ for all their modelled sources.

However, these techniques are complicated by the fact that the fractional 
ionization and 
$\zeta$ are only partly independent of eachother, depending also on other free 
parameters such as the oxygen and carbon depletion, the abundance of the low 
ionization potential metals and the density.
Caselli et al. (\cite{caselliwalmsley}) performed a detailed study of the ionization fraction, 
X(e$^-$), and $\zeta$ in dense cores. They compared simple analytical 
expressions for these quantities (dependent on the two abundance ratios; 
[DCO$^+$]/[HCO$^+$] and [HCO$^+$]/[C$^{18}$O] and the free parameters listed 
above) with the predictions of a more elaborate chemical model, evolved to 
steady-state, and found reasonably good agreement, particularly for $\zeta 
<10^{-17}$s$^{-1}$. In addition, they also noted that the abundance ratios do 
show considerable time-dependence on timescales $<10^5$years. As the sources 
that they have modelled are all quiescent cores (from Butner et al. \cite{butnerlada}) they 
conclude that they are older than 10$^5$years and that chemical equilibrium can 
be assumed. However, the chemical model that they used assumed that the 
physical conditions are fixed. In a dynamically evolving system, the chemical 
timescales may be longer and the initial conditions less well constrained.

van der Tak and van Dishoeck (\cite{vandertakvandishoeck}) also modelled the HCO$^+$ 
and H$_3^+$ abundances (in steady state) on the assumption that the H$_2$O and 
O$_2$ abundances are low (due to freeze out) unless T$>$100K, when mantle 
sublimation occurs. 
They found that the value of $\zeta$ inferred from the H$_3^+$ is 
systematically higher than that inferred from the HCO$^+$ abundances. Noting that HCO+ traces dense gas, they 
speculate that the main cause of this discrepancy is the presence of intervening clouds 
along the line of sight, but again we postulate that a significant cause may be 
the non-equilibrium character of the chemistry.

As described above, in contrast to diffuse clouds, the relevant timescale in
dark clouds is approximately $10^{6}$ years, which may be significantly 
longer than the dynamical timescale in some scenarios. Moreover,
there is increasing evidence to suggest that many of the structures in the 
interstellar medium may be more dynamically active than had been previously 
believed. Thus, for example, Elmegreen (\cite{elmegreen}) and Hartman et al. (\cite{hartmann}) have 
suggested that many of the translucent clumps detected in CO are transient. 
This has prompted plausible models of (MHD) transient structure formation (e.g. 
Falle \& Hartquist \cite{fallehartquist}). These models explain the presence of small, 
non-gravitationally bound clumps, which are found to have lifetimes of $\sim 10^5$ years. 
This timescale is significant as it is less than the time taken for the complete 
conversion of H to H$_2$ and for chemical equilibrium to be established.
The chemical implications of an initially H-rich chemistry were investigated by 
Rawlings et al. (\cite{rawlingshartquist}) who identified the presence of strong chemical 
signatures in such circumstances.
In a cloud of density $\sim 10^4$cm$^{-3}$ the dynamical (free-fall), chemical 
and freeze-out timescales are all comparable ($\sim$few$\times 10^5$ years), so 
that the temporal and spatial variations of molecular abundances are expected 
to be sensitive to the boundary conditions (Rawlings \cite{rawlings}).

We suggest, therefore, that in dynamically evolving regions, the origin 
of the ionization can be ambigious. Specifically, a significant fraction of the ionization that 
is inferred for dark clouds may be a `fossilised' remnant of diffuse cloud 
conditions.
In other words the dark cloud carries chemical signatures of its more diffuse 
nascent state and has not yet established dark cloud chemical equilibrium.
If this is true, then the inferred ionization rates for these objects may be 
incorrect.

\section{CHEMICAL INDICATORS OF $\zeta$ IN DARK CLOUDS}

The chemical networks within dark molecular clouds are well-known and, in their 
early stages, follow a reasonably linear chain of reactions. 
Without recourse to complex chemical models, it is possible to identify the key 
formation and destruction channels for those species which are sensitive to the 
the value of $\zeta$. This simplification allows simple analytical 
expressions which allow the determination of $\zeta$ from a knowledge of the 
abundances of the relevant chemical species to be defined.
We investigate the validity of two well-known analytical expressions for 
$\zeta$;
\begin{enumerate}
\item $\zeta_1$, based on the abundance of H$_3^+$ (see e.g. Lepp, Dalgarno and 
Sternberg \cite{leppdalgarno}). H$_3^+$ is primarily formed by cosmic ray ionization
(\ref{reac1}) and rapid reaction with H$_2$ (\ref{reac2}). In dense
environments it is mainly lost through reactions with O, H$_2$O and CO
((\ref{reac3}), (\ref{reac7}), and (\ref{reac8})). Thus, in equilibrium:
\begin{equation}\label{zeta1}
\zeta_{1}=\frac{n(H_3^+)}{X(H_2)}
[k_{\ref{reac8}}X(CO)+k_{\ref{reac3}}X(O)+k_{\ref{reac7}}X(H_2O)]
\end{equation}
\item $\zeta_2$, based on the HCO$^+$ abundance (see e.g. van der Tak \& van 
Dishoeck \cite{vandertakvandishoeck}). HCO$^+$ is formed by reaction (\ref{reac8}) and primarily lost 
through dissociative recombination (\ref{reac9}) and reaction with water 
(\ref{reac10}). Thus, after substituting for $n({\rm H}_3^+)$ from equation 
(\ref{zeta1}) we obtain;
\begin{eqnarray} \label{zeta2}
\lefteqn{
\zeta_2=\frac{n(HCO^+)\left[k_{\ref{reac7}}X(CO)+k_{\ref{reac3}}X(O)+
k_{\ref{reac7}}X(H_2O)\right]}{k_{\ref{reac8}}X(CO)X(H_2)}}\nonumber\\
 & \qquad \times\left[k_{\ref{reac9}}X(e^-)+k_{\ref{reac10}}X(H_2O)\right]
\end{eqnarray}
\end{enumerate}
In these expressions, the subscripts of the rate coefficients ($k_i$) refer to
the reaction numbers given in the previous section, and the fractional
abundance of a species, $i$, is defined to be the ratio of its number density
to the total hydrogen nucleon density; 
\begin{equation} 
X_i = \frac{n(X)}{n(H)+2n(H_2)}
\end{equation}.

\section{THE MODEL}

The accuracy of both estimators for $\zeta$ was investigated using a 
one-point model of the time-dependent chemistry. The model treats protostellar 
clouds as being of uniform density and temperature, both of which change with 
time in three distinct phases:
\begin{enumerate}
\item {\em Phase I:} Diffuse cloud conditions ($n_0=100$cm$^{-3}$, T=100~K, 
A$_{v0}$=0.5). The system is allowed to evolve until no abundance changes by more than 0.1\% in an 
individual time step. At this point we deem equilibrium to have been reached.
\item {\em Phase II:} Spherically symmetric, homogeneous, free-fall collapse
to a pre-specified terminal density ($n_{max}=10^5$cm$^{-3}$). Initially, 
the collapse is assumed to occur isobarically, so that the temperature is 
inversely proportional to the number density, subject to the constraint that it 
does not fall below some minimum value (T$_{dark}=10$K).
As the collapse is assumed to be both spherically symmetric and homogeneous, 
the extinction $A_v$ is given by

\end{enumerate}

\begin{equation}
A_{v}=A_{v0}\left(\frac{n}{n_0}\right)^{\frac{2}{3}}
\end{equation}

\begin{enumerate}
\setcounter{enumi}{2}

\item {\em Phase III:} Relaxation to chemical equilibrium in dark cloud 
conditions (constant density and temperature, T$_{dark}$=10~K)
\end{enumerate}  

The chemistry includes 87 gas-phase and 36 solid-state chemical species, 
composed of the elements H, He, C, N, O, S and Na and linked through 1265 chemical 
reactions. 
The elemental abundances are given in Table \ref{table:abund}.
These values are broadly representative of cosmic abundances, with a depletion 
factor of 0.5$\times$ included for carbon. (Recent observations (e.g. Meyer, Cardelli \& Sofia 1997) have suggested that nitrogen and oxygen are also depleted by 50\% in diffuse clouds. However, we find that including this extra depletion results in only extremely minor changes to our results.)
 
We have investigated two extremes of metallicity; high (as in Table 
\ref{table:abund}) and low (as in Table \ref{table:abund}, but with the
abundances of S and Na (the representative low ionization potential metal)
reduced by a factor of 0.01$\times$. These values are consistent with those
used by Caselli et al. (\cite{caselliwalmsley}) and others.

\begin{table}
\begin{tabular}{|c|c|}
\hline
Species & Abundance$/ X\left(H\right)$\\
He & 0.1\\
C & 1.87 x $10^{-4}$\\
N & 1.15 x $10^{-4}$\\
O & 6.74 x $10^{-4}$\\
S & 8.00 x $10^{-6}$\\
Na & 2.00 x $10^{-6}$\\
\hline
\end{tabular}
\caption{Elemental abundances by number, relative to hydrogen}
\label{table:abund}
\end{table}

The rate coefficents, $k_i$, for key reactions $i$=2,3,7,8,9 and 10 are taken 
from UMIST ratefiles (Millar et al. \cite{millarrawlings}, \cite{millarfarquhar}) and are given in Table 
\ref{table:rates}. It should be noted, however, that since our study is 
concerned with the comparison of analytical indicators with modelled 
abundances, both of which depend on the same rate coefficients, our results 
are not sensitive to the adopted values of $k_i$.

\begin{table}
\begin{tabular}{|c|c|}
\hline
$k_2$ & 2.08 x $10^{-9}$\\
$k_3$ & 8.0 x $10^{-10}$\\
$k_7$ & 5.9 x $10^{-9}$\\
$k_8$ & 1.7 x $10^{-9}$\\
$k_9$ & 1.1 x $10^{-7}\times (300/T)$\\
$k_{10}$ & 2.5 x $10^{-9}$\\
\hline
\end{tabular}
\caption{Rate coefficients for key reactions (cm$^{3}$s$^{-1}$)}
\label{table:rates}
\end{table}

The passage through regions of intermediate extinction is assumed to be rapid 
and so no attempt has been made to model the details of the photochemistry, 
other than through the normal dependence on A$_v$. 
We do, however, assume that H$_2$ is self-shielding against photodissociation 
at all times, whilst photoionization of C and photodissociation of CO can 
only occur in diffuse cloud (phase I) conditions.

Other than H$_2$ formation (for which we assume k$_{dust}=1.6\times 
10^{-17}$cm$^3$s$^{-1}$, Rawlings et al \cite{rawlingsetal92}), in the standard models we have 
only considered gas-phase chemistry. 
However, in some models we have also allowed gas-phase 
species to freeze out onto the surface of the grains once the extinction rises 
above some critical threshold value (A$_{v,crit.}\sim 3$).
As the sources are dense and cold we do no need to consider the effects of 
surface chemistry as the return of species contained in ice mantles to the gas-phase is unlikely to 
be efficient.

\section{Results}\label{sec:results}
\subsection{Evolution with time}

The evolution of selected species for a typical run is shown in Figure 
\ref{fig:abund}.
This figure clearly shows how the the chemistry starts from primarily atomic
values appropriate for diffuse cloud conditions. As the cloud collapses and 
becomes denser and more opaque to photodissociating radiation, the chemistry
adjusts to a more molecule-rich, lower ionization state characteristic of
dark clouds.

In order to assess the validity of the analytical estimators of $\zeta$, we 
calculate the logarithmic ratio of the rates inferred from the two analytical 
expressions ($\zeta_1$, $\zeta_2$) to the `true' value used in the chemical 
models ($\zeta_{true}$):
\begin{equation}
\rm  R_i = \log_{10}\left(\frac{\zeta_i}{\zeta_{true}}\right)
\end{equation}
and we plot this value as a function of time.
Thus, if $\zeta_i$ were to give a perfectly accurate value of $\zeta$ then 
$R_i$ would be zero.

Results are given in Figure~\ref{zetacrat} for several values of $\zeta_{true}$
($10^{-18}$, $10^{-17}$, $10^{-16}$, $10^{-15}$, and $10^{-14}$s$^{-1}$).

The first thing to note about these results is 
that, at late times (in phase III, once dynamical activity has ceased) the 
curves all converge to values that are very close to 0.
This is rather reassuring in the sense that (although the estimators are based 
on huge simplifications of the chemical networks) the expressions are 
remarkably accurate indicators of the cosmic ray ionization rate for a wide 
range of $\zeta$ {\em when chemical equilibrium and dynamical steady state can 
be assumed}.

The second point to note is that whilst there are clearly very large 
discrepancies during the earlier (collapse) phase, the expressions give accurate results 
almost as soon as the collapse phase is halted, with no notable relaxation 
period.

However, during the collapse (phase II) we can see that significant
discrepancies are present; $\zeta_1$ consistently underestimates 
$\zeta_{true}$, but for the lower ionization rates, this is only by a factor of
$\sim 2-3$. However, if $\zeta_{true}$ were as high as
10$^{-16}-10^{-15}$s$^{-1}$, then the factor is much larger $\sim 10-10^3$.
$\zeta_2$, which is derived incorporating additional reactions in the same
network used to derive $\zeta_1$, is extremely sensitive to the abundances of species
such as H$_2$O, HCO$^+$, and especially CO (See equation \ref{zeta2}). 
These abundances change dramatically during the collapse phase.
The complex structure visible in Figure~\ref{zetacrat}(b) is partly
attributable to this chemical complexity and is partly a result of 
inaccuracies in the numerical integration.
However, as with $\zeta_1$, the magnitudes of the discrepancies are largest in 
the ealiest stages of collapse. But, in contrast to $\zeta_1$, $R_2$ spans
a smaller range ($\sim -1$ to $1$) for the values of $\zeta_{true}$ that were 
investigated. Also we can see that the smaller values of $\zeta_{true}$
($<10^{-16}$s$^{-1}$) are {\it overestimated} by $\zeta_2$, whilst the larger
values of $\zeta_{true}$ are {\it underestimated}. From these results it would 
seem that $\zeta_2$ would always be $\gtappeq 10^{-17}$s$^{-1}$, even though
$\zeta_{true}$ may be somewhat smaller. 

\subsection{Sensitivity to the free parameters}

We have investigated the sensitivity of our results to the various 
free parameters in our model. These include the terminal density ($n_{max}$), 
the terminal temperature (T$_{dark}$), the choice of initial conditions 
corresponding to diffuse cloud
equilibrium or atomic abundances, whether or not freeze-out processes are included, 
the metallicity, and the rate of collapse in phase II.

In models where the final density ($n_{max}$) was varied by up to two orders
of magnitude, $R_1$ was slightly closer to zero for the smaller values of
$\zeta_{true}$, whilst $R_2$ was found to be quite insensitive to $n_{max}$. 

A range of post-collapse temperatures (T$_{dark}$) between 5 and 20 K were 
also investigated for a test case with $\zeta_{true}=10^{-16}$s$^{-1}$. 
The effects on $R_1$ were small ($\sim20\%$) and are restricted to the early 
stages of collapse. In these early stages  $R_1$ is larger (closer to zero) whereas the opposite effect is 
noted for $R_2$. The dependence on T$_{dark}$ is even
more marginal and is also restricted to early times, but here it is the lower
temperatures that result in $R_2$ being closer to zero.

It has been shown (Rawlings et al. \cite{rawlingshartquist}) that the time-dependence of the 
chemistry may be sensitive to the assumed initial conditions - and in
particular the initial H$:$H$_2$ ratio. In our standard model we assume that
the gas is initially characterized by diffuse cloud conditions, with all 
hydrogen in molecular form.
To investigate the possible effects of differing initial conditions on
the validity of the $\zeta$ estimators we have also performed calculations where
the gas is assumed to be intially atomic (H, He, C$^+$, N, O, S$^+$ and Na$^+$) 
with no H$_2$.

We find that $R_2$ is hardly affected by the choice of initial conditions, but 
some differences are noted for $R_1$. Results are shown in Figure 
\ref{zetainit}. For the larger values of $\zeta_{true}$, $\zeta_1$ is larger by a
factor of $\sim 10\times$ - bringing it closer to zero, in the early stages of 
phase II. For smaller values of $\zeta_{true}$ ($<10^{-15}$s$^{-1}$) the differences 
are rather marginal.

Estimates of $\zeta$ and the fractional ionization are closely 
linked, and the latter will be coupled to the assumption that are made 
concerning the metallicity of the gas. Therefore, in addition to our standard 
abundances we have considered a ``low metallicity'' model, in which the 
abundances of the two low ionization potential metals (S and Na) are reduced by 
a factor of 100 to $8\times 10^{-8}$ and $2\times 10^{-8}$ respectively.

The results show a remarkable insensitivity to the metallicity and, as noted 
above, only a relatively small dependence on the initial conditions. 
This helps confirm the hypothesis investigated by this study; that the 
$\zeta$ estimators used are more sensitive to the dynamical status of the observed 
sources than any other individual chemical or physical parameter. 

To simulate accelerated collapse (Lintott et al. \cite{Lintott}) we artificially increased the rate of change 
of the density by a factor of two.
This also had very minimal affects which further supports the idea that the observed  
discrepancies originate from near-instantaneous dynamical effects rather than
the existence of a longer-term chemical hysterisis.
Other factors investigated were the extinction, and the temperature of the 
cloud prior to collapse. Neither had a significant effect.

The result obtained from these parameter variations is
therefore that $R_1$ and $R_2$ are relatively insensitive to the various free
parameters and are primarily determined by the dynamical activity within the
cloud. 

In our standard model we did not include any form of gas-grain interaction, but
we have also investigated the effects of freeze onto the surface of the dust grains by allowing gas-phase
species to freeze-out once A$_v>$3.0~mag. - similar to the critical value of 
the extinction required for the presence of ice water bands in Taurus 
(Whittet et al. \cite{whittetgerakines}). This means that freeze-out commences during the
collapse phase (II). No desorption mechanisms have been included. Results are shown for both $R_1$ and $R_2$ in Figure~\ref{zetafreeze}.

Two clear conclusions can be drawn from this figure: Firstly, if gas-grain processes
are operating - as is likely in the denser regions of the
interstellar medium - then the analytical estimators {\it always} yield 
inaccurate values for $\zeta$, and secondly, the origins of the discrepancies 
(deviations of $R_i$ from 0) are separable. In the collapse phase the 
differences have a dynamical origin, as in the gas-phase models, whilst in
the static, post-collapse phase the differences are entirely a consequence of
the freezing out of molecules onto dust grain surfaces.

\section{Discussion and Conclusions}

Two related methods of estimating the true ionization rate, $\zeta_{true}$,
have been tested for a wide range of parameters. These were based on chemical
networks centered on reactions of H$_3^+$ ($\zeta_2$ using an extended version 
of the network used by $\zeta_1$).
In those cases where gas-grain interactions were ignored, we found that 
chemical equilibrium is rapidly re-established after the end of collapse, and both analytical 
expressions give accurate measures of the true cosmic ray ionization rate in the 
absence of dynamical activity. If freeze-out processes are important, then the
analytical expressions are inaccurate at all times.

There are two components to the modelled discrepancy between the analytical
expressions for $\zeta_1$, $\zeta_2$ and the true cosmic ray ionization rate 
$\zeta_{true}$. Firstly, the fact that 
the analytical expressions ($\zeta_1$ and $\zeta_2$) are simplifications of a 
rather complex chemistry and only make allowance for the dominant chemical 
pathways, and secondly deviations from the assumption of equilibrium.
Of course it is the latter that is the subject of this paper but, as noted in the 
previous section, it is reassuring to see that - at late times - the simple
analytical expressions are reasonably accurate and the contribution from the former
is relatively small. However, this is not always the case and 
has been investigated in some detail by Caselli et al. (\cite{caselliwalmsley}) 
(see their Figure 9) who concluded that, even in an equilibrium approximation, 
the analytical expressions may overestimate $\zeta$ by a factor of $\sim 2-3$ 
for the cases where $\zeta >10^{-17}$s$^{-1}$.

In our models we (obviously) use the same set of rate co-efficients in the 
analytical expressions and the full time-dependent models. Therefore, as 
the results are presented as the ratios of these quantities, they are 
reasonably robust and insensitive to the values of the adopted rate
co-efficients.
 
In all of the cases that we investigated, however, we found that the analytical
expressions are very inaccurate in rapidly evolving, dynamically active 
regions. Discrepancies of several orders of magnitude are present, especially 
for the largest values of $\zeta_{true}$. 
Such conditions may, indeed, be appropriate in many dark clouds that have 
previously been thought of as quiescent entities.
In support of this hypothesis, it should be noted that
there is an increasing body of evidence that suggest that dark molecular clouds 
are not monolithic entities, but consist of ensembles of clumps and 
sub-structure that are not resolved with single-dish molecular line 
observations. For example, populations of distinct sub-cores have been seen in 
TMC-1 Core D (Peng et al. \cite{penglanger}) and L673 (Morata et al. \cite{moratagirart}).  
Moreover, most of these cores are not self-gravitating, but are probably transient 
entities or varying size/mass and dynamical status.
This picture has been supported by recent chemical models (Garrod, Williams \& 
Rawlings \cite{garrodwilliams}) which identify the sub-cores as objects that grow and decay on 
timescales of $\sim 10^6$ years.

Some sensitivity to physical parameters was found, in particular to the density 
reached after collapse. However, these effects were much smaller that the differences 
observed in the same core at different stages of collapse. It can therefore be 
seen that these methods cannot accurately determine the cosmic ray ionization 
rate unless collapse can be definitely shown to have been completed. This forms a challenging observational requirement. 

Geballe et al. 2003 use the observed $\mathrm{H_3^+}$ abundance in several dark clouds to constrain the cosmic ray ionization rate. Oka et al. 2005 compare these measurements of cosmic ray ionization to their measurements of $\zeta$ in diffuse clouds. However, the results presented in this paper suggest that such a comparision is invalid unless the dark clouds being studied were sufficiently old for the dynamical effects revealed by our models to be unimportant.

In their study of massive star-forming regions, van der Tak and van Dishoeck 
(\cite{vandertakvandishoeck}) found that the cosmic ray ionization rate as deduced from H$_3^+$ 
observations was greater than that as deduced from HCO$^+$ observations. That 
is to say, models using $\zeta$ derived from H$_3^+$ observations resulted in a significant over-production of HCO$^+$.
They speculated that, whilst the HCO$^+$ measure probes only the dense 
molecular 
gas, contributions from foreground layers (e.g. in photon dominated regions
or translucent clouds) may be responsible for the discrepancies. This is a valid 
explanation, but we note that dynamical activity may help explain the 
discrepancies; the differences between the values of $\zeta$ inferred from the 
H$_3^+$ and the HCO$^+$ observations (effectively quantified by $R_1-R_2$)
depends on the `true' value of $\zeta$ and the dynamical status of the 
observed source. The more dynamically active the source, the greater the 
discrepancy. 

In conclusion, it is possible that, if a reasonably full dataset of
observations of key molecular tracers is available for any given source,
the absolute values of $\zeta_1$ and $\zeta_2$ and the
difference between $R_1$ and $R_2$ (and perhaps other analytical
expressions for $\zeta$) could be used to simultaneously constrain the true ionization rate 
{\it and} the dynamical status of the source.
If it turns out that very similar values are obtained for $\zeta_1$ and 
$\zeta_2$ then that is an indication that the observed cores are chemically 
`old' and dynamically quiescent.

\begin{acknowledgements}
C.J.L. is supported by a PPARC studentship.
\end{acknowledgements}

\begin{figure}
\includegraphics[width=0.9\textwidth]{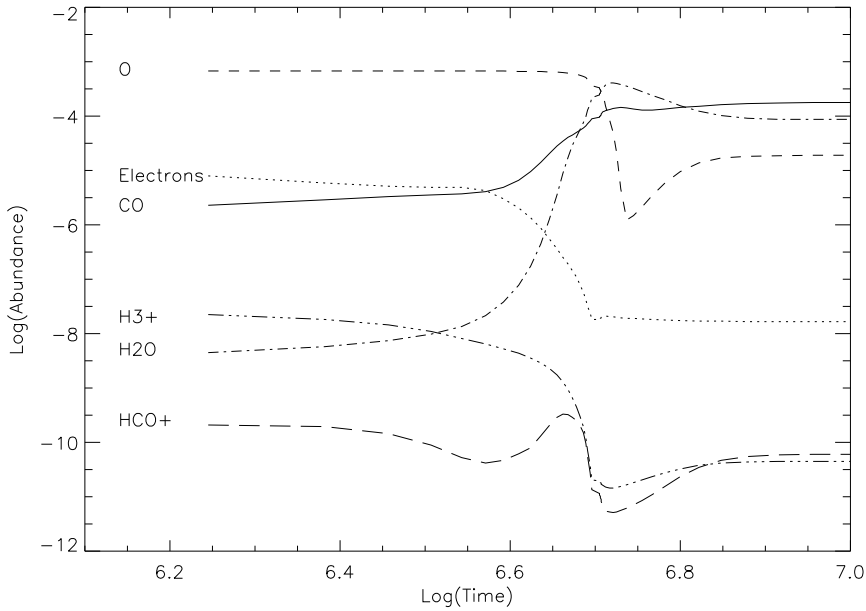}
\caption{The evolution of the fractional abundances, relative to hydrogen,
of key species used in the determination of the cosmic ray ionization rate.
In this model, the cosmic ray ionization rate is 
$\zeta_{true}=1\times 10^{-17}$s$^{-1}$, and the final temperature and density
are 10~K and $10^5$cm$^{-3}$ respectively. Collapse ends approximately at 
Log(time/years)=6.6}
\label{fig:abund}
\end{figure}

\begin{figure}
\includegraphics[width=0.9\textwidth]{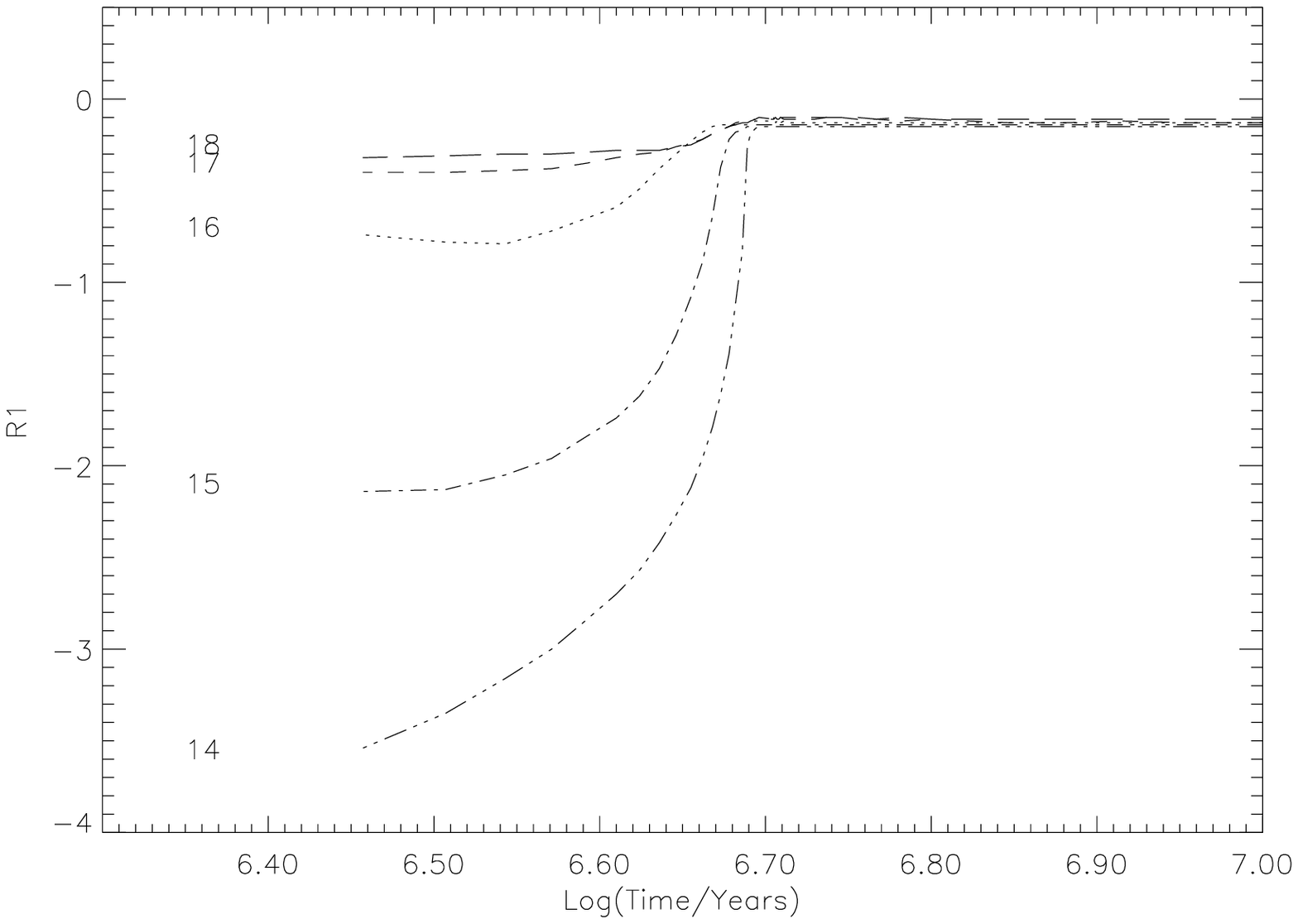}
\includegraphics[width=0.9\textwidth]{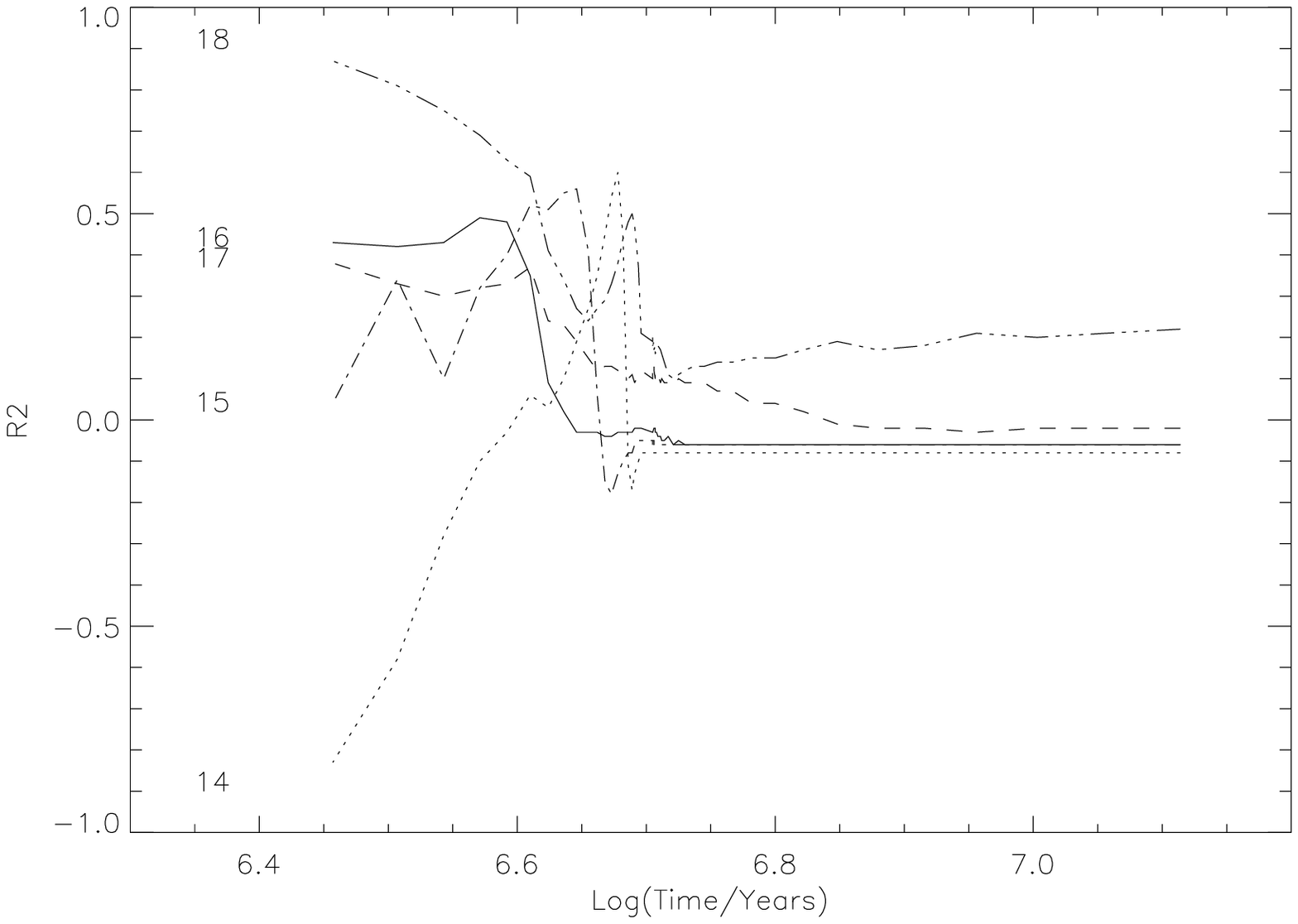}
\caption{The variation of (a) $R_1$, and (b) $R_2$ with time during
phases II (collapse) \& III (post-collapse) for several values of the cosmic
ray ionization rate ($\zeta_{true}$). 
The labels on each curve correspond to -log($\zeta_{true}$). Note that R1 has a much larger range than R2}
\label{zetacrat}
\end{figure}

\begin{figure}
\includegraphics[width=0.9\textwidth]{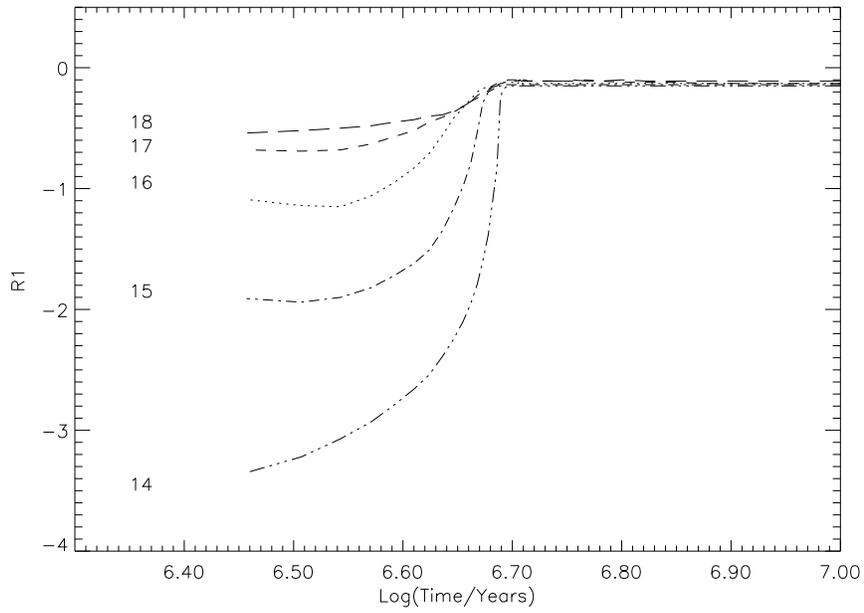}
\caption{Variation of $R_1$ with time for a model in which the chemical initial conditions (Phase I) are characterized by an atomic gas with no H$_2$ present. As before, the labels on each curve correspond to -log($\zeta_{true}$).} 
\label{zetainit}
\end{figure}

\begin{figure}
\includegraphics[width=0.9\textwidth]{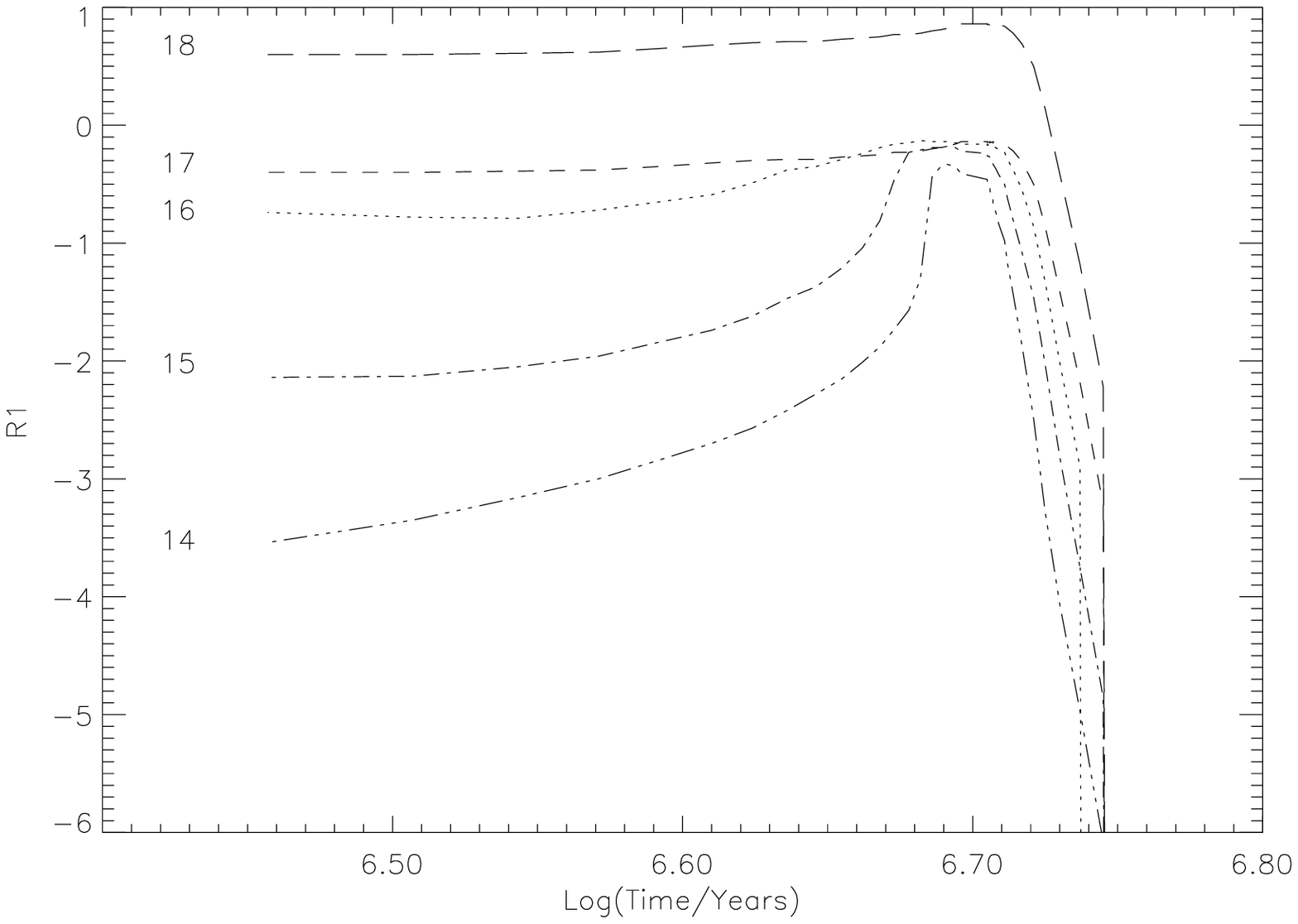}
\includegraphics[width=0.9\textwidth]{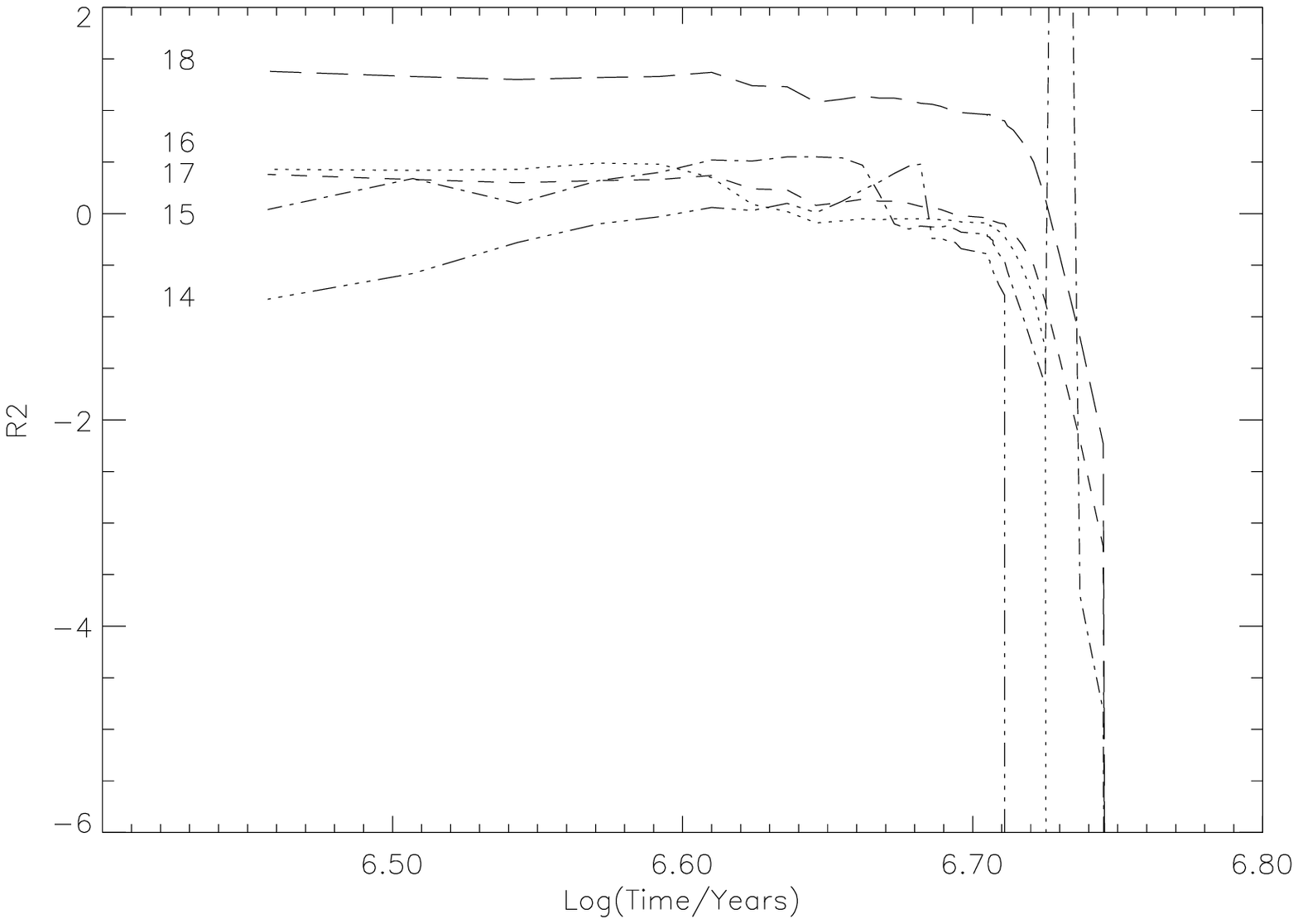}
\caption{Variation of (a) $R_1$ and (b) $R_2$ with time for a model in which 
freeze-out occurs once $A_v>3.0$~mag. As before, the labels on each curve correspond to -log($\zeta_{true}$), and the same
line style has been used for each $\zeta_{true}$ in figure 4(b). The large jumpin R2 at Log(time/years)=6.74 is an effect of
numerical integration.} 
\label{zetafreeze}
\end{figure}

\label{lastpage}

\end{document}